\documentclass[12pt]{article}
\usepackage{graphics}
\usepackage{epsfig}
\usepackage{latexsym}
\newcommand{\beq}{\begin{equation}}
\newcommand{\eeq}{\end{equation}}
\newcommand{\beqa}{\begin{eqnarray}}
\newcommand{\eeqa}{\end{eqnarray}}
\newcommand{\beqar}{\begin{eqnarray*}}
\newcommand{\eeqar}{\end{eqnarray*}}

\newcommand{\tn}{\tilde{n}}

\newcommand{\eg}{{\it e.g.,}\ }
\newcommand{\ie}{{\it i.e.,}\ }
\newcommand{\labell}[1]{\label{#1}} %{\label{#1}\qquad_{#1}} %
\newcommand{\reef}[1]{(\ref{#1})}

\newcommand\veps{\varepsilon}

\newcommand\bg{{\bar g}}
\newcommand\bn{{\bar\nabla}}
\def\IR{{\hbox{{\rm I}\kern-.2em\hbox{\rm R}}}}
\begin{document}

\thispagestyle{empty}
\rightline{\small hep-th/9908175 \hfill McGill/99-18}
\vspace*{2cm}

\begin{center}
{\bf \LARGE Spin-Two Glueballs, Positive Energy Theorems }\\[.25em]
{\bf \LARGE and the AdS/CFT Correspondence}
\vspace*{1cm}

Neil R. Constable\footnote{E-mail: constabl@hep.physics.mcgill.ca} and
 Robert C. Myers\footnote{E-mail: rcm@hep.physics.mcgill.ca}\\
\vspace*{0.2cm}
{\it Department of Physics, McGill University}\\
{\it Montr\'eal, QC, H3A 2T8, Canada}\\
\vspace{2cm} ABSTRACT
\end{center}
We determine the spectrum of graviton excitations in the background
geometry of the AdS soliton in $p$+2 dimensions. Via the
AdS/CFT correspondence, this corresponds to determining
the spectrum of spin-two excitations in the dual effective
$p$-dimensional field theories. For the cases of D3- and M5-branes,
these are the spin-two glueballs of QCD$_3$ and QCD$_4$, respectively.
For all values of $p$, we find an exact degeneracy of the spectra
of these tensor states and certain scalar excitations. 
Our results also extend the perturbative proof of a positive energy
conjecture for asymptotically locally AdS spacetimes (originally proposed for
$p$=3) to an arbitrary number of dimensions.
 
\vfill \setcounter{page}{0} \setcounter{footnote}{0}
\newpage

\section{Introduction} \label{intro}                 

The AdS/CFT correspondence~\cite{jaun,gkp,ed} --- for a
comprehensive review, see ref.~\cite{review} --- has provided new
perspectives on the holographic principle \cite{thooft}, which asserts
that a consistent theory of quantum gravity in $d$ dimensions must have
an alternate formulation in terms of a nongravitational theory
in $d-1$ dimensions. This equivalence is implemented in the AdS/CFT
correspondence with a duality between a gravitational theory in
$d$-dimensional anti-de Sitter space and a conformal field theory 
on a $(d-1)$-dimensional ``boundary'' space. Using this correspondence,
one can gain new insights into both of the theories on either side of
the duality. On the one hand, quantum gravity is reformulated as an
ordinary quantum field theory and so one has a new framework with
which to study the perplexing puzzles surrounding black holes and
Hawking evaporation. On the other hand, the duality
has become a useful tool with which to study a wide class of 
strongly coupled field theories in a non-perturbative framework. 

In the latter context, the AdS/CFT correspondence has been
extensively used to study ``QCD-like'' field theories by considering
a variety of solutions to AdS supergravity ---
see, for example, refs.~\cite{edd,ooguri,mello,more,mini,zero,scaler}.
The original proposal by Witten\cite{edd} was made in the setting
of string theory, where the conformal field theory is 
a supersymmetric gauge theory. One can describe
ordinary (\ie nonsupersymmetric) Yang-Mills theory by compactifying
one of the spatial directions on a ``small'' circle and imposing
antiperiodic boundary conditions on the fermions around this direction.
In this case, the additional fermions and scalars appearing in the
supersymmetric field theory would acquire large masses 
of the order of the compactification scale, leaving the gauge fields
as the only low energy degrees of freedom.
On the supergravity side, this proposal corresponds to considering
spacetimes which are asymptotically  AdS locally, but not globally.
The relevant supergravity solution was found\cite{edd} to be
the ``AdS soliton'' --- following the nomenclature of ref.~\cite{positive}
--- that is the double analytic continuation of a planar AdS black hole.
In this geometry, the circle direction smoothly contracts to a point
in the interior and as a result of this nontrivial topology, the
supergravity fermions (and their dual gauge theory counterparts) are
antiperiodic on the asymptotic circle.

In the special case of supergravity on AdS$_5$, which
arises in the throat geometry of D3-branes, the dual field theory
is four-dimensional ${\cal N}=4$ $SU(N)$ super-Yang-Mills theory and so upon
compactification, the above construction produces an effective model
of three-dimensional ordinary Yang-Mills theory, \ie QCD$_3$~\cite{edd}.
Similarly beginning with AdS$_7$, which arises from M5-branes, and compactifying
two directions, one produces a model of QCD$_4$~\cite{edd}.
This idea has been exploited\cite{ed,ooguri} to make predictions about the 
large-N behavior of these QCD theories at strong coupling. Specifically, 
it was shown\cite{edd} that these theories generically produce an area law
for spatial Wilson Loops and
possess a mass gap, both of which are evidence for confinement. The spectrum
of scalar glueballs for QCD$_{3,4}$ was been calculated\cite{ooguri}, and
found to be remarkably similar to that calculated by lattice
techniques\cite{QCD3,QCD4,latt}. A similar scalar spectrum was calculated
for arbitrary D$p$-branes in ref.~\cite{mini}.

In this paper we will extend the work of refs.~\cite{ooguri,mini}
by calculating the complete spectrum of spin-two excitations of the
effectively $p$-dimensional field theories which are dual to the 
AdS soliton in $p+2$ dimensions. In the cases,
for which $p=3$ or $p=5$, this amounts to a determination of the 
spin-two glueball spectra in QCD$_3$ and QCD$_4$. For all values of
$p$, we find a surprising degeneracy between the spin-two states and scalar
excitations associated with a minimally coupled massless scalar field propagating
on the AdS soliton background.

Another interesting development motivated by the AdS/CFT correspondence was 
the proposal of a new positive energy conjecture for asymptotically locally
AdS spaces~\cite{positive}. This investigation was originally motivated
by the observation that in asymptotically flat spacetimes
geometries in which an asymptotic circle is contractible are
unstable because the energy can be arbitrarily negative\cite{nega}.
For gravity with a negative cosmological constant, one does find that the
AdS soliton has a finite negative energy. However, in ref.~\cite{positive}
for five dimensions, it was shown that this solution is perturbatively stable
and that the negative energy is naturally identified with the Casimir energy
of the dual field theory --- see also ref.~\cite{stress}.
Hence motivated by the AdS/CFT correspondence,  
it was proposed that the AdS soliton is in fact the minimum energy 
solution with the given asymptotic structure. 
Our present calculation extend the results of ref.~\cite{positive} by
providing a perturbative proof of the stability of the AdS soliton which holds
in all dimensions $d\ge 4$. Hence we are lead to extend the positive
energy conjecture of ref.~\cite{positive} beyond five dimensions, which
was the focus of their discussion.

The paper is organized as follows: In section 2, we review the standard
approach to calculating glueball spectra using AdS/CFT correspondence.
In section 3, we calculate the 
spectrum of spin-two excitations in the field theory
dual to the AdS soliton in arbitrary dimensions. In
section 4, we consider our results as a perturbative proof of the positive 
energy conjecture of ref.~\cite{positive} extended to arbitrary
dimensions. Section 5 contains a further discussion of  our results.

\section{Review of Scalar Spectra} \label{review}

In the framework of string theory where the AdS/CFT duality is best understood,
simple backgrounds are both supersymmetric and conformally invariant.
Both of these symmetries must be eliminated to construct models of
real world QCD. Witten's suggestion\cite{edd} described above breaks
supersymmetry with the antiperiodic fermion boundary conditions, and
breaks the conformal invariance by introducing a scale, namely the
size of the circle. The appropriate supergravity background satisfying
the desired boundary conditions is the AdS soliton.
For this solution, Witten argued that the
supergravity fields must have a discrete spectrum, and hence
a mass gap would exist in the dual field theory\cite{edd}.
The AdS/CFT correspondence gives the interpretation that the 
the supergravity modes represent excitations of particular
gauge theory operators, which then possess a discrete
mass spectrum. These results were verified in detail in
refs.~\cite{ooguri,mello}.
 
Generalizing to $p+2$ dimensions, the AdS soliton metric may be written as
\beqa
ds^2 &=& \frac{r^2}{L^2}\left(f(r)d\tau^2 + 
\eta_{\mu\nu}dx^{\mu}dx^{\nu}\right)+
\frac{L^2}{r^2}f^{-1}(r)dr^2   
\nonumber
\\
&&{\rm with}\ f(r) = \left(1-\frac{R^{p+1}}{r^{p+1}}\right)
\labell{eq:metric}
\eeqa
where $\eta_{\mu\nu}dx^{\mu}dx^{\nu}$ is the $p$-dimensional 
Minkowski metric. This geometry can be constructed by a double analytic
continuation of a planar AdS black hole in horospheric coordinates 
--- see, for example, the discussion in ref.~\cite{positive}.
The geometry is only locally asymptotically AdS because implicitly
the $\tau$ coordinate is chosen to be periodic in order to
avoid a conical singularity at $r=R$. Choosing the period to be
\beq
\beta = \frac{4\pi L^2}{(p+1)R}
\labell{period}
\eeq
the circle parametrized by $\tau$ smoothly shrinks to a point
at $r=R$.  As a result, in the context of supergravity, the fermionic
fields and their dual gauge theory counterparts are antiperiodic
around the $\tau$ circle.

This geometry was considered for the special cases $p=3$ and $5$ in
ref.~\cite{ooguri,mello} where the mass spectra for $0^{++}$ glueballs in
QCD$_3$ and QCD$_4$ were calculated.\footnote{The
discussion of $p=5$ is rephrased there in terms of the throat geometry
of D4-branes, using the duality between type IIa superstrings
in ten dimensions and M-theory in eleven dimensions \cite{atom}.}
Ref.~\cite{ooguri} also presents the mass spectra for $0^{--}$ 
and $0^{-+}$ glueballs by considering the appropriate supergravity fields.
Scalar glueball spectra have also been analyzed in detail in 
refs.~\cite{sfet3}.

\begin{figure}[ht!]
\center{\includegraphics{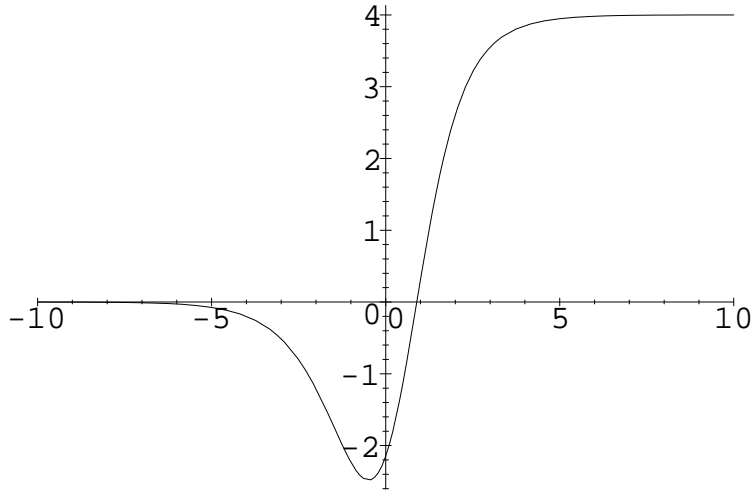}}
\caption{Plot of the effective potential $V(y)$ in eq.~\reef{eq:pot} 
for $p=3$ and $\frac{M^2L^4}{R^2}=50$. The figure demonstrates 
the existence of two classical turning points.} 
\label{fig:ttpot}
\end{figure}

One proceeds by solving the linearized wave equation for a minimally
coupled massless
scalar field propagating in the above background \reef{eq:metric}.
Specifically one considers the equation
\beq
\Box \phi = \frac{1}{\sqrt{-g}}\partial_{\mu}\left(\sqrt{-g}g^{\mu \nu}
\partial_{\nu}\,\phi\right) = 0 \ .
\labell{eq:kgeqn}
\eeq
In the case of $p=3$, \ie D3-branes, this scalar corresponds to the
dilaton which is dual to the gauge theory operator which is the
supersymmetric extension of Tr$F^2$
\cite{igor,gkp,ooguri}. One introduces the ansatz $\phi=b(r)e^{ik\cdot x}$,
where $b(r)$ is the radial profile to be determined and the momentum $k^\mu$
is a $p$-vector with only components in the Minkowski space directions
spanned by the coordinates $x^\mu$. One then has $k^2=-M^2$ giving the 
invariant mass-squared of the dual scalar operator in the effective
$p$-dimensional field theory. Substituting
this ansatz into eq.~\reef{eq:kgeqn} one obtains an ordinary differential
equation for the radial profile $b(r)$,
\beq
\frac{\partial^2 b(r)}{\partial r^2} + \frac{(p+2)r^{p+1}-R^{p+1}}
{r\left(r^{p+1}-R^{p+1}\right)}\frac{\partial b(r)}{\partial r} 
+\frac{M^2L^2r^{p-2}}{r\left(r^{p+1}-R^{p+1}\right)}b(r) = 0
\labell{ode}
\eeq
To facilitate our analysis, this equation can be put into a 
Schr\"odinger-like form~\cite{mini,sfet3} by redefining the wave function $b(r)$
as $b(r)=\beta (r) \chi (r)$ where,
\beq
\beta (r) = \sqrt{\frac{r-R}{r(r^{p+1}-R^{p+1})}}.
\labell{redef}
\eeq
and then performing a change of variables according to $r=R(1+e^y)$.
Eq.~\reef{ode} now takes the form:
\beq
-\chi^{\prime \prime}(y) + V(y)\,\chi(y)  =  0
\labell{schro}
\eeq
where the effective potential is given by
\beqa
V(y) & =& \frac{1}{4} +\frac{e^{2y}\left(
p(p+2)\left(1+e^y\right)^{2(p+1)}
-2p(p+2)\left(1+e^y\right)^{p+1}-1\right)}{4\left(1+e^y\right)^2\left(
\left(1+e^y\right)^{p+1}-1\right)^2}
\nonumber
\\
&&\ -\frac{M^2L^4}{R^2}\frac{e^{2y}\left(1+e^y\right)^{p-3}
}{\left(1+e^y\right)^{p+1}-1}\ \ .
\labell{eq:pot}
\eeqa
This complicated analytic form for the potential belies a relatively
simple shape, as shown in figure \ref{fig:ttpot}. Now one must tune
the potential by adjusting the parameter $M^2$ so that eq.~\reef{schro}
yields a normalizable bound state at {\it zero energy.} One thing that is
easily verified is that the scalar field fluctuations produce no instabilities
for this supergravity background. That is there are no tachyonic solutions
since when $M^2$ is negative (or even positive but small --- see below)
the potential well in figure \ref{fig:ttpot} disappears and $V(y)$ is
everywhere positive.
%nnn for the wish list: can we overlay with a dotted line the potential
%nnn with M^2<0 -- this request has zero^+ priority
%r maybe later...it is probably worth doing though.
The equation \reef{schro}
can be solved either numerically\cite{ooguri}, by matching asymptotic
solutions\cite{mello} or in the 
WKB approximation\cite{mini,sfet3}\footnote{Note that the calculation
described here for the AdS soliton
in $p$+2 dimensions is {\it not} the same as that in ref.~\cite{mini}.
There Minahan considers the propagation of the dilaton 
for a generalized D$p$-brane background. Thus the results only 
agree with the present analysis for the special case $p=3$.}.
We will discuss here solving the spectrum for these scalar excitations
in WKB approximation outlined in refs.~\cite{mini,sfet3}.
The potential given in eq.~\reef{eq:pot} 
takes the following asymptotic forms
\beqa
V(y \gg 0)&=&{(p+1)^2\over4}-\frac{p(p+2)}{2}e^{-y}
+\left({3p(p+2)\over4}-{M^2L^4\over R^2}\right)e^{-2y}+\ldots
\nonumber
\\
V(y\ll 0)&=&\left({p+2\over4}-\frac{M^2L^4}{(p+1)R^2}\right)e^y
+\ldots
\labell{asympot1}
\eeqa
Hence the classical turning point at large $y$ is approximately
\beq
y_{+} = \log{\frac{2ML^2}{(p+1)R}}
\labell{turnout}
\eeq
where this result is valid to $O(M^{-1})$ in an expansion with
${R\over ML^2}\ll 1$. The inner turning point is located at
\beq
y_{-} = -\infty\ .
\label{eq:turn1}
\eeq
as long as $M^2L^4/R^2>(p+1)(p+2)/4$, which is consistent with
the previous assumption. If the latter inequality is not satisfied,
the potential is in fact everywhere positive and no bound states exist
even without making the WKB approximation.
In terms of the original $r$ coordinate in eq.~\reef{eq:metric},
these turning points correspond to,
\beq
r_{+}=R+\frac{2}{p+1}ML^2
\qquad{\rm and}\qquad
r_{-}=R\ .
\labell{turnR}
\eeq
In the WKB approximation then, one finds a zero-energy bound state for
\beq
\left(n-\frac{1}{2}\right)\pi = \int_{y_{-}}^{y_{+}}\sqrt{V(y)}dy 
\labell{eq:WKB}
\eeq
with $n$ being a positive integer.
Following refs.~\cite{mini,sfet3}, we expand the integral above
as a series in powers of $\frac{R}{ML^2}$ and consider only terms appearing at
${ O}(M)$ and ${ O}(M^0)$. The corrections at the next order will appear at 
${O}(M^{-1})$. Adding up the contributions and solving for $M$, one finds
\beq
M^2(p) = n\left(n+\frac{p-1}{2}\right)\frac{16\pi^3}{\beta^2}
\left(\frac{\Gamma\left(\frac{p+3
}{2(p+1)}\right)}{\Gamma\left(\frac{1}{p+1}\right)}\right)^2 + O(n^0)
\labell{eq:ttspec}
\eeq
where $\beta$ is the period of the compact
coordinate $\tau$ given in eq.~\reef{period}.
For $p=3$ and $5$, which are relevant for QCD$_{3,4}$, the above 
expression gives\footnote{In comparing with the results of 
ref.~\cite{mini}, note that $M^2($here$)=\pi^2M^2($there$)$.}
\beqa
M^2(p=3) & \simeq &\frac{56.67}{\beta^2}n\left(n+1\right) +O(n^0)
\nonumber
\\
M^2(p=5) & \simeq &\frac{29.36}{\beta^2}n\left(n+2\right) +O(n^0)
\labell{ttnum}
\eeqa

\section{The Graviton Spectrum}

The purpose of the present paper is to repeat these calculations
for gravitons propagating in the AdS$_{p+2}$ soliton,
and hence to calculate the spectrum of spin-two
excitations for the effective $p$-dimensional field theories.
The latter excitations are those associated with the stress-energy tensor
$T_{\mu \nu}$ of the conformal field theory as this is the operator coupling
to the AdS metric perturbations, \ie the gravitons, according to the AdS/CFT
correspondence \cite{kleb,kleb2} --- see discussion in ref.~\cite{stress}.
To determine the spectrum, we must solve the equations of
motion for the gravitons on the background \reef{eq:metric}.
Specifically we write the perturbed metric as
\beq
g_{ab} = \bg_{ab} +h_{ab}
\labell{pert}
\eeq
where $\bg_{ab}$ denotes the background metric~\reef{eq:metric}
which is a solution of Einstein's equations in $p+2$ dimensions
with a negative cosmological constant:
\beq
R_{ab}+{p+1\over L^2}g_{ab}=0\ .
\labell{einst}
\eeq
Linearizing the latter equations of motion, the metric perturbation
$h_{a b}$ must satisfy~\cite{wald}:
\beq   
\frac{1}{2}\nabla_a\nabla_b\,h^c{}_c +\frac{1}{2}\nabla^2h_{a b} - 
\nabla^c\nabla_{(a}h_{b) c}-{p+1\over L^2}h_{ab}=0
\labell{eq:lingrav}
\eeq
where the background metric $\bg_{ab}$ is used to raise and lower indices 
as well as to define the covariant derivatives in this equation.
Now in analogy to the scalar field calculations,
we make the ansatz that the
graviton is given by $h_{ab} = H_{a b}(r)e^{ik\cdot x}$ where $H_{a b}(r)$ is
the radial profile tensor and $k^\mu$ is a $p$-dimensional momentum vector
with $k^2=-M^2$. To determine the spectrum, we must solve eq.~\reef{eq:lingrav}
with this ansatz as an eigenvalue problem for the mass $M$. 
 
A feature here is the fact that $H_{a b}(r)$
is a tensor and there are thus a variety of polarizations, or graviton modes, 
that need to be considered. Further there are the ambiguities
in the metric perturbations arising from diffeomorphism invariance, which
we handle by imposing a ``transverse gauge'': $H_{a\mu}k^\mu = 0$.
For massive excitations 
we may always, via the appropriate Lorentz boost, choose to 
work in the rest frame\footnote{Our analysis was done for the completely
general case, however, here we anticipate a 
spectrum with positive definite mass squared.} so that the momentum can be
written as $k^\mu=\omega\,\delta^\mu_{t}$. In this case, the transversality condition
becomes,
\beq
H_{a\mu}k^\mu = 0 \;\;\;\Rightarrow \;\;\;H_{a t}=0 \;\;\;\;\forall a
\labell{trans}
\eeq
Our implicit notation for the $p$-dimensional Minkowski space coordinates is
$x^\mu=(t,x^i)$ with $i=1,\ldots,p-1$.

\subsection{Simple Transverse Traceless Polarizations}

In presenting the solutions of eq.~\reef{eq:lingrav}, we begin
with the gravitons polarized in the directions parallel to the hypersurface
spanned by the coordinates $x^\mu$, \ie we give the solutions
with
\beq
H_{\tau a}  =  H_{r a}  =  0 = H_{a \mu}k^\mu \qquad\forall\, a\ ,
\labell{pol1}
\eeq
where we have included the gauge condition \reef{trans} in this
list of restrictions.
From the point of view of the dual field theory, these excitations
correspond to spin-two states. Other polarizations would
be interpreted as scalar or vector states in the $p$-dimensional
field theory --- see below.

A consistent solution is provided by
the following ansatz
\beq
H_{ab}=\varepsilon_{ab}\,{r^2\over L^2}H(r)
\labell{ttpol}
\eeq
where the constant polarization tensor $\varepsilon_{ab}$ satisfies the
restrictions given in eq.~\reef{pol1}. Solving the equations of motion
\reef{eq:lingrav} imposes one further restriction on the polarization,
namely, it must be traceless
\beq
\eta^{\mu\nu}\varepsilon_{\mu\nu}=0\ .
\labell{traceless}
\eeq
Thus eq.~\reef{ttpol} describes $(p+1)(p-2)/2$ independent modes, which can
be described as $(p-1)(p-2)/2$ off-diagonal polarizations, \eg
\beq
\varepsilon_{12}=\varepsilon_{21}=1\ ,\qquad 
{\rm otherwise}\ \varepsilon_{ab}=0
\labell{exampoff}
\eeq
and $(p-2)$ traceless diagonal polarizations, \eg
\beq
\varepsilon_{11}=-\varepsilon_{22}=1\ ,\qquad 
{\rm otherwise}\ \varepsilon_{ab}=0\ .
\labell{exampdia}
\eeq
We also observe that $(p+1)(p-2)/2$ corresponds to precisely the number
of polarizations of a massive spin-two particle in $p$ dimensions.

For all of these independent polarizations, the radial profile $H(r)$
satisfies the same differential equation. Substituting the above ansatz
\reef{ttpol} into eq.~\reef{eq:lingrav} yields
\beq
\frac{\partial^2 H(r)}{\partial r^2} + \frac{(p+2)r^{p+1} - R^{p+1}}{r(r^{p+1}
 - R^{p+1})}\frac{\partial H(r)}{\partial r} + \frac{M^2L^4r^{p-3}}
 {\left(r^{p+1}-R^{p+1}\right)}H(r) = 0
\labell{ttode}
\eeq
As for the case of the scalar equation \reef{ode},  we
put this equation into
a Schr\"odinger-like form by setting $H(r)=\alpha(r)
\phi(r)$ with
\beq
\alpha (r)= \sqrt{\frac{r-R}{r(r^{p+1}-R^{p+1})}}\ ,
\labell{ttdef}
\eeq
and by again changing variables according to $r=R(1+e^y)$. Eq.~\reef{ttode}
 now takes the form,
\beq
-\phi^{\prime \prime}(y) + V(y)\phi(y)  =  0
\labell{ttschro}
\eeq
where the effective potential is given by,
\beqa
V(y) & =& \frac{1}{4} +\frac{e^{2y}\left(
p(p+2)\left(1+e^y\right)^{2(p+1)}
-2p(p+2)\left(1+e^y\right)^{p+1}-1\right)}{4\left(1+e^y\right)^2\left(
\left(1+e^y\right)^{p+1}-1\right)^2}
\nonumber
\\
&&\ -\frac{M^2L^4}{R^2}\frac{e^{2y}\left(1+e^y\right)^{p-3}
}{\left(1+e^y\right)^{p+1}-1}\ \ .
\labell{eq:ttpot}
\eeqa

Now let us compare this effective potential (\ref{eq:ttpot}) with
eq.~\reef{eq:pot} for the minimally coupled massless scalar field
considered in the previous section. One sees immediately
that the effective potentials~\reef{eq:pot} and~\reef{eq:ttpot} are identical!
In fact, the equations of motion \reef{ode} and \reef{ttode} are also
identical. In other words, the scalar and the transverse traceless gravitons
considered here have exactly the same equations of motion and hence
they have identical
mass spectra, for all values of $p$. Via the AdS/CFT correspondence, this
implies the degeneracy of tensor and scalar excitations in the
corresponding dual field theories, \ie
\beq
\frac{M_{2^{++}}}{M_{0^{++}}}=1\ .
\labell{ratio}
\eeq
In the case of QCD$_3$, the prediction is that the glueball spectra
of the operators $TrF^2$ and $T_{\mu \nu}$ are degenerate.
We will comment more on this degeneracy in the discussion section.

Within the WKB approximation, we may read off the spectrum of the transverse
traceless gravitons from the results of the scalar field analysis
in eqs.~\reef{eq:ttspec} and \reef{ttnum}.
Note, however, that the scalar-tensor degeneracy is 
an exact statement, which holds outside of any approximation scheme used
to compute the masses.

\subsection{Exotic Polarizations}

The remaining linearized graviton solutions come in two categories.
The first would appear as vectors in the dual field theory, and they
may be given with the same ansatz as in eq.~\reef{ttpol}
\beq
H_{ab}=\varepsilon_{ab}\,{r^2\over L^2}H(r)
\labell{eq:eppol}
\eeq
however, in this case the nonvanishing components of the polarization tensor take
the form
\beq
\varepsilon_{\tau \mu}=\varepsilon_{\mu\tau}=v_\mu\ ,
\qquad{\rm with}\ k\cdot v=0\ {\rm and}\ v\cdot v=1\ .
\labell{vector}
\eeq
Thus this vector solution contains $(p-1)$ independent modes.
Substituting into the equations of motion \reef{eq:lingrav}
now yields the radial equation
\beq
\frac{\partial^2 H(r)}{\partial r^2} + \frac{(p+2)}{r}
\frac{\partial H(r)}{\partial r} + \frac{M^2L^4r^{p-3}}
 {\left(r^{p+1}-R^{p+1}\right)}H(r) = 0\ .
\labell{vecode}
\eeq
Using
the field redefinition $H(r)=\alpha (r)\phi(r)$ where now $\alpha 
(r)=\sqrt{\frac{r-R}{r^{p+2}}}$
 and again changing variables to the $y$ coordinate as before, we find a
 Schr\"odinger-like equation where the potential is given by:
\beq
V(y) = \frac{1}{4} +\frac{p(p+2)e^{2y}}{4\left(1+e^y\right)^2} -
\frac{M^2L^4}{R^2}
\frac{e^{2y}\left(1+e^y\right)^{p-3}}{\left(1+e^y\right)^{p+1}-1}
\labell{eq:eepot}
\eeq

\begin{figure}[ht!]
\center{\includegraphics{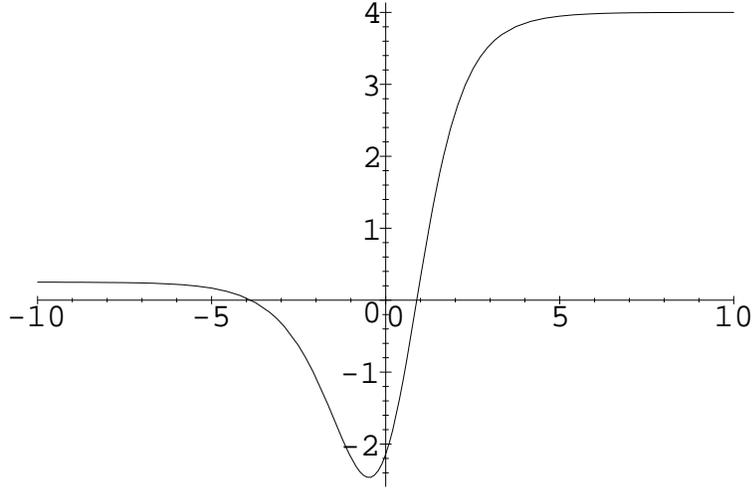}}
\caption{Plot of the effective potential $V(y)$ in eq.~\reef{eq:eepot} 
for $p=3$ and $\frac{M^2L^4}{R^2}=50$. There are two classical 
turning points for { finite} values of $y$.}
\label{fig:semi}
\end{figure}
This potential has the following asymptotic forms,
\beqa
V(y\gg 0)&=& \frac{(p+1)^2}{4}-\frac{p(p+2)}{2}e^{-y}+\left(\frac{p(p+2)}{4}
-\frac{M^2L^4}{R^2}\right)e^{-2y}+\ldots
\nonumber \\
V(y\ll 0)&=&\frac{1}{4}-\frac{M^2L^4}{R^2}e^y+\ldots
\labell{exasymp}
\eeqa
and thus has classical turning points located at:
\beqa
y_{+}&=& \log{\frac{2ML^2}{(p+1)R}}
\nonumber
\\
y_{-}&=& -\log{\frac{4M^2L^4}{R^2(p+1)}}
\labell{eq:turn2}
\eeqa
Note in this case that the inner turning point, $r_{-}$
(given in terms of the original $r$ coordinate),
is situated away from the surface $r=R$. Given these results,
the WKB mass spectrum is:
\beq
M^2(p) =n\left(n+\frac{p+5}{4}\right)\frac{16\pi^3}{\beta^2}
\left(\frac{\Gamma\left
(\frac{3+p}{2(p+1)}\right)}{\Gamma\left(\frac{1}{p+1}\right)}\right)^2
+ O(n^0)\ .
\labell{eq:exspec}
\eeq
Note that this WKB result is very similar to that for the scalars
(and transverse traceless gravitons) given in eq.~\reef{eq:ttspec},
with the only difference being in the coefficient of the $O(n)$ term.
For the cases of interest for QCD$_{3,4}$, this expression yields
\beqa
M^2(p=3) & \simeq &\frac{56.67}{\beta^2}n\left(n+2\right) + O(n^0)
\nonumber
\\
M^2(p=5) & \simeq &\frac{65.86}{\beta^2}n\left(n+\frac{5}{2}\right) + O(n^0)
\labell{seminum}
\eeqa
and so these vector excitations are slightly heavier than the tensors or
scalars. The latter applies generally for $p<7$, while in fact the
vectors become lighter for $p>7$. At $p=7$, one finds that the 
vector spectrum here is degenerate with that for the scalars, however,
this is a degeneracy which only holds within the approximations of our
WKB calculation. Since the effective potentials \reef{eq:pot} and
\reef{eq:eepot} remain different for $p=7$, this degeneracy should be lifted
by higher order corrections, but we can not say which set of states will
give the heavier spectrum.

The last polarization is diagonal and would appear as a scalar 
excitation in the $p$-dimensional field theory.
To construct this final solution of the linearized equations
of motion \reef{eq:lingrav}, we begin with the following ansatz:
\beqa
H_{\tau \tau}(r)&=& -{r^2\over L^2}f(r)H(r) 
\nonumber
\\
H_{\mu\nu}(r)&=& {r^2\over L^2}\left(\eta_{\mu\nu} a(r)+
{k_\mu k_\nu\over M^2}(b(r)+a(r))\right)
\nonumber
\\
H_{rr}(r)&=& \frac{L^2}{r^2}f^{-1}(r)c(r)
\nonumber
\\
H_{r\mu}(r)&=& ik_\mu\,d(r)
\labell{guess}
\eeqa
where $f(r)$ is as defined in eq.~\reef{eq:metric} and all other components
of $H_{ab}$ vanish. Note that we have not imposed the
transverse gauge condition \reef{trans} in the ansatz above.
Substituting this ansatz into eq.~\reef{eq:lingrav}, consistency
of the equations requires
\beqa
a(r)&=&\frac{1}{p-1}H(r) 
\nonumber
\\
c(r)&=& \frac{(p+1)R^{p+1}}{2pr^{p+1}-(p-1)R^{p+1}}H(r)
\labell{consistent}
\eeqa
while the function $d(r)$ is determined by $b(r)$ and $H(r)$ through the
relation
\beqa
d(r)=-\frac{r^2}{2L^2M^2}\frac{\partial b(r)}
{\partial r}+\frac{(p+1)r^{2}R^{p+1}}{2\left(2pr^{p+1}-(p-1)R^{p+1}
\right)L^2M^2}\frac{\partial H(r)}
{\partial r}
\nonumber \\
+\frac{p(p+1)^2r^{p+2}R^{p+1}H(r)}{\left(2pr^{p+1}-(p-1)R^{p+1}\right)
^2L^2M^2}\ .
\labell{jjj}
\eeqa
Hence one has the freedom to choose $d(r)$ by making the appropriate choice 
for the functional form of $b(r)$. For the present calculations, a convenient
choice is to eliminate the derivative terms appearing in eq.~\reef{jjj}, which
allows one to express the polarization tensor explicitly in terms of $H(r)$.
This gauge choice is accomplished by choosing
\beq
b(r)=-\frac{(p+1)R^{p+1}}{2pr^{p+1}-(p-1)R^{p+1}}H(r)
\labell{chooz}
\eeq
which yields
\beq
d(r)=\frac{2p(p+1)^2r^{p+2}R^{p+1}H(r)}{M^2L^2\left(2pr^{p+1}-
(p-1)R^{p+1}\right)^2}\ .
\labell{chooz2}
\eeq

An alternative choice is
to set $b(r)=-a(r)$, for which the diagonal components of the polarization 
along the $x^\mu$ directions have the form $\varepsilon_{\mu\nu}\propto
\eta_{\mu\nu}$. 
Thus this choice makes manifest the Lorentz invariance of
this polarization in the $p$ dimensions of the field theory --- otherwise
it would only be Lorentz invariant up to a gauge transformation ---
and so it is obvious that these modes are actually dual to
scalar excitations in the field theory.\footnote{Note that with $b=-a$
in eq.~\reef{guess}, there are still the off-diagonal components $H_{r\mu}\propto k_\mu$,
but they do not
constitute an independent vector. Further one could remove these components with
an infinitesimal diffeomorphism with $v_a=-d(r)\exp(ik\cdot x)\,\delta^r_a$, but then
the ansatz for the diagonal components becomes even more complicated.} 

Using the results in eq.~\reef{consistent}, the coupled set of 
equations~\reef{eq:lingrav} reduce to a single second order linear
ODE for the function $H(r)$ which upon redefining $H(r)$ by 
$H(r)=\eta (r)\psi (r)$ where
\beq 
\eta (r)=\sqrt{\frac{r-R}{r(r^{p+1}-R^{p+1})}}
\labell{redef2}
\eeq
and changing to the $y$ coordinate of the previous sections takes the 
standard Schr\"odinger form \reef{schro} with a potential given by
\beqa
V(y) & =& \frac{1}{4} +\frac{e^{2y}\left(
p(p+2)\left(1+e^y\right)^{2(p+1)}
-2p(p+2)\left(1+e^y\right)^{p+1}-1\right)}{4\left(1+e^y\right)^2\left(
\left(1+e^y\right)^{p+1}-1\right)^2}
\nonumber
\\
&&\ -\frac{M^2L^4}{R^2}\frac{e^{2y}\left(1+e^y\right)^{p-3}
}{\left(1+e^y\right)^{p+1}-1}
-{2(p-1)(p+1)^3\left(1+e^y\right)^{p-1}e^{2y}\over
\left(\left(1+e^y\right)^{p+1}-1\right)
\left(2p\left(1+e^y\right)^{p+1}-p+1\right)^2}
\ \ .
\labell{eq:diagpot}
\eeqa
The potential is written here in a way so that it only differs from the
scalar potential \reef{eq:pot} by the addition of the last term.

\begin{figure}[ht!]
\center{\includegraphics{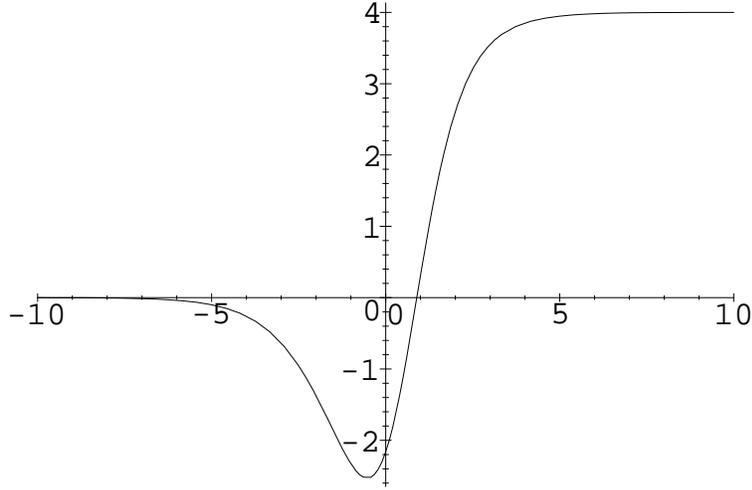}}
\caption{Plot of the effective potential $V(y)$ in eq.~\reef{eq:diagpot} 
for $p=3$ and $\frac{M^2L^4}{R^2}=50$. Note the similarity 
to fig.~\reef{fig:ttpot}.} 
\label{fig:diagpot}
\end{figure}
In the asymptotic regions, the potential has the following forms
\beqa
V(y\gg 0)&=& \frac{(p+1)^2}{4}-\frac{p(p+2)}{2}e^{-y}+\left(\frac{3p(p+2)}{4}
-\frac{M^2L^4}{R^2}\right)e^{-2y}+\ldots
\nonumber \\
V(y\ll 0)&=&-\left((7p-10)+\frac{M^2L^4}{(p+1)^2R^2}\right)e^y
+\ldots
\labell{asympot}
\eeqa
As can be seen in eq.~\reef{asympot}, the behavior for large positive $y$ is 
identical to that given in eq.~\reef{asympot1} thus the turning point 
is given by $y_+$ in 
eq.~\reef{turnout}. However, the behavior for large negative $y$ is
slightly different but there is still a turning point at $y_{-}=-\infty$.
Thus, to leading order in the ${1}/{M}$ expansion\cite{mini} of the WKB
integral~\reef{eq:WKB}, the mass
spectrum is identical to that given in eq.~\reef{eq:ttspec},
\beq
M^2 = n\left(n+\frac{p-1}{2}\right)\frac{16\pi^3}{\beta^2}\left(\frac{\Gamma
\left(\frac{3+p}{2(p+1)}\right)}{\Gamma\left(\frac{1}{p+1}\right)}\right)^2
+ O(n^0)
\labell{skall}
\eeq
The WKB mass spectra for these glueballs in QCD$_{3,4}$ is given by 
eq.~\reef{ttnum}. However, this degeneracy between the present
scalar states and those in section 2 holds only within the WKB
approximation. It will be lifted by higher order corrections.

\section{Positive Energy Conjecture}

In this section, we consider the implications of the above results for
the positive energy conjecture put forth in ref.~\cite{positive}.
While the conjecture there was formulated for five dimensions, here we
wish to consider the generalization to an arbitrary number of dimensions
(greater than three). Hence consider solutions of Einstein's equations
in $p+2$ dimensions with a negative cosmological constant, which
asymptotically approach the AdS soliton metric $\bg_{ab}$ given
in eq.~\reef{eq:metric} with
\beq
g_{ab}=\bg_{ab}+h_{ab}
\labell{conjecm}
\eeq
where the perturbation 
$h_{ab}$ has the following asymptotic behavior
\beq
h_{ab}=O(r^{-p+1}), \;\;\;h_{a r}=O(r^{-p-1}), \;\;\;
h_{rr}=O(r^{-p-3}) \qquad {\rm with}\ a , b \neq r\ .
\labell{eq:asymp}
\eeq
Then $E(g_{ab})\geq E(\bg_{ab})$ with equality obtained if and only if
$g_{ab}=\bg_{ab}$. In ref.~\cite{positive}, it was verified 
that, for the case of $p=3$, the AdS soliton is 
perturbatively stable to all linearized deformations of the metric. 
In the following, we apply directly the 
results obtained in section 3 to give a  proof of perturbative stability
of this solution in arbitrary numbers of dimensions.  

The essential point is that in the spectra calculated in the previous
section, we found that $M^2>0$ in all cases. Hence through the AdS/CFT
correspondence, these metric fluctuations will be dual to positive energy
excitations of the field theory, with $E=\sum n_i\,\omega_i$ where $n_i$ is the number of
quanta excited in a particular mode with frequency $\omega=k^t$.
From this point of view, negative energies or instabilities
would be signaled by the appearance of ``tachyonic'' excitations with
$M^2<0$. For a complete discussion of perturbative stability of the AdS
soliton, one should also consider all linearized solutions including those
with momenta in the compact $\tau$ direction. Naively, one expects that
adding such internal momenta will only increase the masses. In fact, this
intuition is correct, and we leave the detailed analysis of these modes
to Appendix A. Hence $M^2>0$ for all graviton modes, and so 
this indicates that the AdS soliton is stable against all linearized
metric fluctuations in any number of dimensions.

Given that we wish to determine whether a particular solution of
Einstein's equations is perturbatively stable against metric deformations,
we should be able to address this question in a purely gravitational framework,
without reference to the AdS/CFT correspondence. We do so now by
considering the construction for gravitational energy by Abbot and Deser\cite{Abbot}.
One begins by dividing the metric as in eq.~\reef{conjecm}, but now the metric
deformation $h_{ab}$ is considered to be defined globally. Given this decomposition
and the fact that the background metric $\bg_{ab}$ satisfies Einstein's equations
with a negative cosmological constant,
\beq
\bar{R}_{ab}-{1\over2}\bg_{ab}\bar{R} - {p(p+1)\over2L^2}  \bg_{ab}=0\ ,
\labell{unstein}
\eeq
the equations determining the deformation $h_{ab}$ may be written as\cite{Abbot}:
\beq
R^L_{ab}-\frac{1}{2}\bg_{ab}R^L+{p(p+1)\over2L^2} h_{ab}=T_{ab}\ .
\labell{stressdef}
\eeq
Here the superscript $L$ denotes that these curvature terms have been
linearized in $h_{ab}$, while the ``stress-energy tensor'' on the right-hand
side is defined to include all terms second order and higher in $h_{ab}$.
Because the left-hand side of eq.~\reef{stressdef} obeys the background Bianchi
identity, ${\bar\nabla}^a\left(R^L_{ab}-{1/2}\bg_{ab}R^L+p(p+1)/(2L^2) h_{ab}
\right)=0$, the Einstein equations \reef{stressdef} dictate
that
\beq
{\bar\nabla}^aT_{ab}=0
\labell{exacti}
\eeq
as an exact result, where $\bn_a$ denotes the covariant derivative
with respect to the background metric $\bg_{ab}$. Further,
it is understood that all indices are raised and lowered here using the background
metric. Now, given a Killing vector $\xi^a$ of the background metric, one has
\beq
{\bar\nabla}^a\left(T_{ab}\xi^b\right)=0
\labell{exactj}
\eeq
as a result of the Killing vector identity, $\bn_a\xi_b+\bn_b\xi_a=0$, and the
symmetry of $T_{ab}$. Denoting the one-form
\beq
\left(T\cdot\xi\right)=dx^a\,T_{ab}\xi^b\ ,
\labell{onef}
\eeq
then the dual $(p+1)$-form is closed, \ie
\beq
d *\left(T\cdot\xi\right)=0\ ,
\labell{closed}
\eeq
given the vanishing divergence in eq.~\reef{exactj}. Thus integrating this
form over a $p+1$ dimensional surface with a time-like normal
\beq
\int *\left(T\cdot\xi\right)
\labell{conserf}
\eeq
yields a conserved charge.\footnote{Note that this integral is
finite given the rate at which $h_{ab}$ vanishes at asymptotic infinity in
eq.~\reef{eq:asymp}.} Now focusing our attention on the time-like
Killing vector $\xi^a\partial_a=\partial_t$ of the AdS soliton background,
the Killing energy is defined as the conserved charge
\beq
E=\frac{1}{8\pi G}\int *\left(T\cdot\xi\right)
\labell{endef}
\eeq
where the normalization is chosen here to match that
of alternate definitions for asymptotically AdS spacetimes, such as in
refs.~\cite{stress,HH,vijay,abhay}.

The preceding discussion refers to an exact solution of Einstein's equations, \ie
given the metric decomposition \reef{conjecm} with the background satisfying
eq.~\reef{unstein}, eq.~\reef{stressdef} is satisfied to all orders in $h_{ab}$.
In the analysis of the previous section, we have only been considering metric
deformations which satisfy the linearized Einstein equations \reef{eq:lingrav},
\ie all of the higher order terms implicit in $T_{ab}$ in eq.~\reef{stressdef}
have been dropped. Consistent with this linearized analysis, 
our strategy will be to identify the stress-energy density of
our perturbed spacetimes to be the contributions quadratic in $h_{ab}$
in eq.~\reef{endef}. This approach will be sufficient to identify whether
or not the background metric is {\it perturbatively} stable.

It is simplest to apply this analysis to the transverse traceless modes of
section 3.1. We write the metric perturbation as
\beq
h_{ab} = \varepsilon_{ab}{r^2\over L^2} H(r,\tau,x^\mu)
\labell{extend}
\eeq
where the constant polarization tensor satisfies the constraints
$\veps_{ra}=\veps_{\tau a}=0=\veps_{a \mu}\nabla^\mu H$, as well as
$\eta^{\mu\nu}\veps_{\mu\nu}=0$.
Also the polarization will be normalized such that $\varepsilon^{ab}\varepsilon_{ab}=2$,
as in the examples in section 3.
Inserting this ansatz into eq.~\reef{endef} and keeping only the
quadratic terms, one finds
\beq
E=\frac{1}{32\pi G}\int drd\tau\, d^{p-1}\!x\,\left(\frac{r}{L}\right)^{p-2}
\left((\partial_tH)^2-H\partial^2_tH\right)
\labell{simenergy}
\eeq
where we integrated by parts in both $r$ and $\tau$, and used the linearized equations
of motion to produce this simple expression. Substituting for eq.~\reef{extend}
the real part of the solutions found in secion 3.1 or appendix A,
\ie $H(r,\tau,t,x^i)=H(r)\,\cos(k\cdot x+q\tau)$, yields
\beq
E=\frac{\omega^2}{32\pi G}\int drd\tau\, d^{p-1}\!x\,\left(\frac{r}{L}\right)^{p-2}
H(r)^2
\labell{sttenergy}
\eeq
where $\omega=k^t$.
Clearly now, this expression is manifestly positive definite, and so the AdS soliton
is perturbatively stable to deformations by these modes. Note that this classical
result is consistent with that stated above from the AdS/CFT correspondence.
In particular, the Lorentz-invariant measure for momentum modes
is $d^{p-1}k/\omega(\vec k)$ which is why the field theory energy is proportional to 
$\omega$ while
the classical result \reef{sttenergy} has a factor of $\omega^2$. Further the quantum
number operator in the field theory will be proportional to the field squared as
in eq.~\reef{sttenergy}, and the factor of $G^{-1}$ would be absorbed using
the canonical normalization for the quantum graviton field --- see, \eg ref.~\cite{velt}.
Note that here we assume a simple local mapping between the perturbative quantum
field in the supergravity theory, \ie the gravitons, and the dual quantum field
in the conformal field theory, as in the discussions of ref.~\cite{perturb}
--- see below, however.
Performing the same analysis for the exotic polarizations of section 3.2
and appendix A is more complicated. The results are simple, however, if we consider
only the modes which are independent of the compact coordinate $\tau$. 
For example, consider our ansatz for the vector modes 
\beq
h_{ab} = \varepsilon_{ab}{r^2\over L^2} H(r)\,\cos k\cdot x
\labell{extend2}
\eeq
where the nonvanishing components of the polarization tensor are
$\veps_{\tau \mu}=\veps_{\mu\tau}=v_\mu$ with $k\cdot v=0$ and $v\cdot v=1$.
Note that in order to produce a sensible energy, we have again taken the real part
of the ansatz used in section 3. Such a metric perturbation then yields
\beq
E=\frac{\omega^2}{32\pi G}\int drd\tau\, d^{p-1}\!x\,\left(\frac{r}{L}\right)^{p-2}
f^{-1}(r)\,H(r)^2
\labell{vecenergy}
\eeq
where $f(r)$ is defined in eq.~\reef{eq:metric}, and $\omega=k^t$ as above.
Similarly if we take the real part of the ansatz presented in eq.~\reef{qguess}
for the scalar modes,
we find a relatively simple expression for the quadratic Killing energy
\beq
E={p\over p-1}\,\frac{\omega^2}{64\pi G}\int drd\tau\, d^{p-1}\!x\,\left(\frac{r}{L}
\right)^{p-2}H(r)^2\ .
\labell{diagenergy}
\eeq
Both of these expressions are manifestly positive definite, and so obviously the
AdS soliton is perturbatively stable to deformations by these modes.

The gravitational calculation of the energy of the vector and scalar modes
becomes far more complicated. For example, for the vector modes with
$\tau$-dependence, one works with the real part of the ansatz described in 
eq.~\reef{qpol} of the appendix. Inserting this into the above formulae
\reef{endef} yields an extremely long expression for the energy density. One can 
arrange this expression in powers of the AdS radius $L$ giving,
\beq
E=\frac{M^2}{64\pi G}\int drd\tau\,d^{p-1}\!x\,\left(\frac{r}{L}\right)^{p-2}
\left\lbrace\frac
{r^{p+1}\left(1+q^2/M^2\right)^2}{r^{p+1}\left(1+q^2/M^2\right)-R^{p+1}}
H(r)^2 +O\left(\frac{q^2}{L^4}\right)\right\rbrace\ ,
\labell{exenergy}
\eeq
where the remaining implicit terms at $O(q^2/L^{4})$ and $O(q^4/L^{8})$ fill another
page and a half, and are not particularly illuminating. In particular, while one can see
immediately that the first term above is positive
definite, the sign of the remaining higher order contributions is not
obvious. Therefore establishing that the energy in eq.~\reef{exenergy} is
positive definite requires a more extended analysis. We will not present
such an analysis here, but one can be confident in the result since there
is also the much simpler proof relying on the AdS/CFT duality described above.
One also runs into
similar complications in calculating the gravitational energy for the scalar
modes with nontrivial $\tau$-dependence.

As a final point on the discussion of the gravitational energy, it may be disconcerting
that the expressions for the vector and scalar modes with nontrivial
$\tau$-dependence, \eg eq.~\reef{exenergy}, are so complicated. These expressions
suggest that there is a complicated nonlocal relationship between the quantum
operators producing the graviton excitations in the AdS soliton background
and the dual
operators in the field theory. This result may be surprising given that
in ref.~\cite{perturb}, a simple identity was established between 
the AdS and CFT quantum modes for all excitations.
However, that analysis applied for a (super)gravity background which was purely
anti-de Sitter space, and the identification relied on the
large symmetry group of this background. Only a small part of this symmetry group
survives for the AdS soliton background,
and so perhaps it not so surprising that the relationships can become more complicated.
These complications are probably also related to the nonlocal expressions that arise
for various metric components in the relevant graviton modes, \eg see eq.~\reef{ggg}.

\section{Discussion}

In this paper we have calculated the complete spectrum of metric fluctuations
of the $(p+2)$-dimensional AdS soliton given in eq.~\reef{eq:metric}.
The dual conformal field theory will be formulated on a $(p+1)$-dimensional
background geometry $M^p\times S^1$ with metric
\beq
ds^2=\eta_{\mu\nu}dx^\mu dx^\nu +d\tau^2
\labell{bacc}
\eeq
where $\tau$ inherits the same period $\beta$ as in eq.~\reef{period}. In the limit,
that this circle direction is small, the field theory should become effectively
$p$-dimensional. From string theory, it is in this situation that the AdS soliton
has been suggested to give a dual description of ordinary Yang-Mills theory  in
three or four dimensions. For the polarizations which
are confined to the $x^\mu$ directions then, our results correspond 
to the mass spectrum of the spin-two excitations of the effective $p$-dimensional
field theory. A second interesting implication of our analysis is that
since we have found that $M^2>0$ for all of the modes, our results confirm the
perturbative stability of the $(p$+2)-dimensional AdS soliton, and thus
provide evidence the positive energy conjecture of ref.~\cite{positive}
can be extended to arbitrary numbers of dimensions. 

Being a massless
spin-two field, the graviton propagating in any $(p$+2)-dimensional spacetime has
${1\over2}p(p+1)-1$ physical degrees of freedom. In the present case, these
degrees of freedom are organized as representations of the 
$p$-dimensional Lorentz symmetry
of the AdS soliton background. These are: $(p+1)(p-2)/2$ modes in massive
spin-two representations, $p-1$ polarizations in massive vector representations,
and 1 set of massive scalar states. Each of these sets of states is further
labeled by the integer $\tilde{n}$ giving the momentum in the $\tau$-direction,
\ie $q=2\pi \tilde{n}/\beta$. These graviton modes are dual
to operators in the corresponding conformal field theory with the same
quantum numbers. In fact, one knows that the relevant operator is the
stress-energy tensor $T_{ab}$ of the $(p$+1)-dimensional field theory
as this is the operator coupling
to the AdS metric perturbations according to the AdS/CFT
correspondence \cite{kleb,kleb2}.
Of course, the various components of $T_{ab}$ are decomposed according to
the $SO(1,p-1)$ symmetry of the boundary manifold \reef{bacc} matching the
decomposition of the metric fluctuations described above.

Thus when the various graviton modes are excited in the AdS soliton
background, there should be a corresponding excitation in the expectation
value of the stress-energy tensor in the dual field theory description.
One can verify that the metric fluctuations fall off with precisely the appropriate
rate to yield a nonvanishing $\langle T_{ab}\rangle$, \ie $h_{ab}=O(r^{-p+1})$
in $p$+2 dimensions \cite{stress}. Examining the asymptotic behavior of the
effective potential for any of the modes, one finds
\beq
V(y>>0)\simeq {(p+1)^2\over4}+\ldots
\labell{assss}
\eeq
Thus to leading order in this regime, the normalizable solution of the
Schr\"odinger equation behaves as
\beq
\psi(y>>0)\simeq e^{-{p+1\over2}y} \simeq \left({R\over r}\right)^{p+1\over2}\ .
\labell{wavass}
\eeq
Now while the details of the field redefinition factor varies from mode to
mode, one finds that they all have the same asymptotic behavior at large $r$,
namely $\eta(r>>0)\simeq \left({R/ r}\right)^{p+1\over2}$, and hence to leading
order, the radial profile becomes
\beq
H(r)=\eta(r)\psi(r)\simeq \left({R\over r}\right)^{p+1}\ .
\labell{morasss}
\eeq
Finally various components of the metric fluctuations contain $r^2H(r)=
O(r^{-p+1})$, giving precisely the desired fall off.

So for example, the transverse traceless gravitons correspond to
inducing an excitation of the stress-energy in the field theory with
\beq
\langle T_{\mu\nu}\rangle\, \propto \veps_{\mu\nu}\cos(k\cdot x+q\tau)
\labell{expect1}
\eeq
where $\veps$ is the polarization tensor appearing in the metric fluctuation
which satisfies $\veps_{\mu\nu}k^\nu=0=\eta^{\mu\nu}\veps_{\mu\nu}$.
Similarly for the vector modes, one finds from eq.~\reef{qpol} that the nonvanishing
stress-energy components are
\beq
\langle T_{\tau\mu}\rangle \propto v_\mu \cos(k\cdot x+q\tau)
\quad{\rm and}\quad 
\langle T_{\mu\nu}\rangle \propto {q\over M^2}(k_\mu v_\nu+v_\mu k_\nu)
\cos(k\cdot x+q\tau)
\labell{expect2}
\eeq
where $k\cdot v=0$ and $v\cdot v=1$. Here, although there are nontrivial metric
components $h_{r\mu}$, asymptotically they fall off too rapidly to contribute in the
calculation of $\langle T_{ab}\rangle$ \cite{stress}. Finally for the scalar
mode \reef{guess} with $q=0$, one finds
\beq
\langle T_{\tau\tau}\rangle \propto -(p-1) \cos(k\cdot x)
\quad{\rm and}\quad 
\langle T_{\mu\nu}\rangle \propto \left(\eta_{\mu\nu}+{k_\mu k_\nu\over M^2}\right)
\cos(k\cdot x)
\labell{expect3}
\eeq
Note that all of these stresses are traceless and transverse in the $p$+1
dimensional background \reef{bacc}
of the field theory, \ie $\langle T^a{}_a\rangle=0=\langle T_{ab}\rangle k^b$
where $k^a=(k^\mu,q)$. Note that the three expressions above correspond
to the additional stress-energy associated with the graviton. The AdS soliton
itself also induces an nontrivial Casimir stress-energy \cite{stress}.

Note that with a single mode excited in its rest frame,
\ie with $k^a=(M,0,\ldots,0,q)$, no energy density is induced. There are
only stresses in the spatial directions for all such modes.
However, in general there is a nontrivial expectation
$\langle T_{tt}\rangle$, and in fact there are regions of negative as well as positive
energy density because of the cosine factors.
Note however that the negative and positive contributions precisely cancel
out to yield zero total energy at this order. For example, in the cases
where $q\ne0$, integrating over the compact $\tau$-direction is enough to
produce a vanishing
result. This is the expected result since the induced energy densities here are
linear in the amplitude of the metric fluctuations, and we found in section
4 that the total energy was in fact quadratic in this amplitude. Similar observations
were made in ref.~\cite{plt} where the scattering of gravitons in AdS space was
considered. The vanishing total energy emerging from these calculations is a result
of the fact that we have only considered the linearized Einstein equations
in discussing the present graviton modes. If one were to completely solve the full
Einstein equations or at least solve them to next order in the amplitude of the waves,
one would find that the long-range metric perturbations would receive second-order
contributions yielding $E=\int d\tau d^{p-1}\!x^i\,\langle T_{tt}\rangle>0$.
The integrand of the energy
expressions in section 4 would be closely related to the source terms for the second
order contribution in the metric fluctuation $h_{tt}$. These second order
contributions would ensure that while there are locally regions with negative
energy density, they are always accompanied by regions with a larger net positive
energy density, in keeping with the quantum interest conjecture\cite{ford}. 

Since the vector modes couple to $\tau$-components of
the stress-energy tensor (even with $q=0$) in the field theory, as shown in
eq.~\reef{expect2},
and the background metric \reef{bacc} is diagonal, one would require
that these modes should decouple if the AdS soliton is to provide a good
description
of an effective $p$-dimensional field theory.\footnote{The same should be
true for the scalar modes. However, with $q=0$ only
$\langle T_{\tau\tau}\rangle\not=0$ in eq.~\reef{expect3}, and this term
could arise from a contribution proportional to the background metric
\reef{bacc} which is diagonal.}
In fact, however, there is no evidence of such decoupling\footnote{In fact, we already
see that the theory really fails to become $p$-dimensional as the lightest excitations
have masses of the order of the compactification scale.}
as the masses of all states are roughly of the order $\beta^{-1}$. In fact, as noted
before, the vector modes are lighter than the spin-two modes for $p>7$.
This is similar to the non-decoupling of excitations on the internal $S^5$ found
in considering the case of $p=3$ \cite{cousin}. In this string theory context,
one might consider the effect of $\alpha'$-corrections to the string theory action
\cite{standard}. Such higher derivative terms would modify both the 
AdS soliton background \cite{igoralpha} and the equations of motion.
However, we did not pursue these lengthy calculations. In the context of the higher
$S^5$ harmonics, it was found that the leading $\alpha'$-corrections did not produce
the desired decoupling, and it was conjectured that decoupling may only
result through nonperturbative effects \cite{cousin}.

A surprising result that we have found is that for all values of $p$ the spectrum
of the spin-two modes is exactly degenerate with that of a minimally coupled massless
scalar field. In the dual field theory, there will be an exact degeneracy of
the corresponding spin-two and spin-zero excitations. In particular then, in the
case of $p=3$ where the effective field theory is ordinary (\ie nonsupersymmetric)
Yang-Mills theory in three dimensions\cite{edd},
there will be a degeneracy of the spin-two
glueballs associated with the stress tensor $T_{\mu\nu}$, and the spin-zero
glueballs associated with the operator Tr$F^2$. We emphasize that
the degeneracy of these spectra is a consequence of the fact that the
dilaton and the gravitons satisfy precisely the same equation
of motion, and is not an artifact of the WKB calculations presented in
section 3. There is a similar degeneracy
for spin-two and scalar glueballs in QCD$_4$, which is associated with the
AdS$_7$ soliton with $p=5$ \cite{edd} --- see below. This is a surprising prediction of
the AdS/CFT correspondence as current lattice simulations\cite{QCD3,QCD4,latt}
certainly do not provide any evidence of such a remarkable degeneracy.
Rather the lattice calculations suggest that the spin-two excitations should
be more massive than the scalar glueballs. For the lowest lying glueballs,
simulations for QCD$_3$~\cite{QCD3} estimate ${M_{2^{++}}}/{M_{0^{++}}}\simeq 
1.65 \pm 0.04$ for the large $N$ limit.
Similarly, for QCD$_4$~\cite{QCD4}, lattice calculations suggest
${M_{2^{++}}}/{M_{0^{++}}}\simeq 1.39\pm0.13$ in the large $N$ limit.
(The errors quoted are the statistical errors arising in
extrapolating to the continuum from the lattice.) Of course, this discrepancy
between the lattice calculations and the AdS/CFT correspondence is not a direct 
contradiction. The gravity calculations performed here would be valid for
$SU(N)$ gauge theory in the large $N$ limit in of (very) large 't Hooft coupling $g^2N$.
In contrast the lattice results would hold for weak 't Hooft coupling.
Hence it might be that this degeneracy is a result of the strong coupling limit
relevant for the supergravity calculations. Again, it would be interesting
to see to what extent the degeneracy is lifted by $\alpha'$-corrections to the low
energy action. It would also be interesting to see
if this degeneracy survives in other supergravity models exhibiting QCD-like
behavior\cite{zero,scaler} (such as
running couplings and asymptotic freedom).

We have no insight as to why the spectrum of the spin-two graviton modes
should be degenerate with that of a scalar field in general. However, for $p=3$,
it can be related to previous observations about D3-branes. In ref.~\cite{kleb},
it was found that in the background geometry of a D3-brane, the
gravitons polarized along the world-volume had an action of the form,
\beq
I = -\frac{1}{64\pi G_{10}}\int d^{10}x\sqrt{-g}\partial_{\mu}h_{ab}
\partial^{\mu}h_{ab}
\labell{gravact}
\eeq
and thus, from the point of view of the {\it transverse} space and time, 
they behave as minimally coupled massless scalars. As a result, it was shown that
the classical absorption cross-sections were identical for these gravitons and 
the dilaton. Now it is easy to verify that the same observations still hold true
in the geometry of an analytically continued near-extremal D3-brane, as long as
the graviton polarizations have no $\tau$-components. The throat geometry
of such a D3-brane would be the direct product of the AdS$_5$ soliton 
with a five-sphere.
The AdS$_5$ soliton geometry is comprised of the world-volume directions
and the radial direction in the transverse space. In the spectrum calculations
on the AdS soliton background, one focuses essentially on the time and radius part
of the wave equation, which would be identical to that arising in the propagation of
an S-wave in the throat geometry. Thus one can interpret the result, that
the field equations governing the spin-two modes are the same as for a scalar field,
as a reflection of the above observation that these spin-two modes behave
as scalar fields in the transverse space.

In ref.~\cite{mini}, Minahan calculated the spectrum of the dilaton (a minimally
coupled massless scalar) propagating in the throat geometry of an analytically
continued near-extremal D$p$-brane for general $p$. These throat geometries are
not precisely the same as the AdS$_p$ soliton as these backgrounds also contain
a nontrivial dilaton field, except for $p=3$. One may note, however, that there
is a matching of his dilaton spectrum for the D1,3,4-branes with that of a scalar
in the AdS soliton with $p=2,3,5$, respectively. Of course, this is not a surprise
for the D3-brane where the throat geometry is just the AdS soliton with $p=3$.
For the D4-brane, the matching arises because this brane in the Type IIa string
theory can be lifted to an M5-brane in D=11 supergravity \cite{town}, where
the throat geometry contains an AdS$_7$ soliton factor with $p=5$. 
Similarly through a chain of dualities, the D1-brane of the Type IIb string
theory can be related to the M2-brane, where the AdS$_2$ soliton with $p=2$ appears
in the throat geometry. However, note that there are no scalar fields in
D=11 supergravity. The string theory dilaton arises as part of the
eleven-dimensional metric. Hence the matching of the spectra that we noted
above only occurs because of the degeneracy of spin-two graviton
modes and scalar field excitations in the AdS soliton background, since Minahan
\cite{mini} actually calculated the spectrum of a particular mode of the graviton in
a dual description. In fact, in the case of the M2-brane for which there are no
spin-two excitations, the degeneracy arises as an artifact of the WKB approximation
for which the spin-zero graviton spectrum \reef{skall} matched that of
the scalar field \reef{eq:ttspec}.

The case of the M5-brane is of particular interest as it may give a
dual description of QCD$_4$ \cite{edd}.
One can regard the field theory as being obtained from the
compactification of the six-dimensional $(0,2)$ conformal field theory
on $S^1 \times S^1$
where the first circle is much smaller than the second circle,
on which one also imposes supersymmetry breaking boundary conditions \cite{edd}.
In terms of the AdS/CFT correspondence, this is equivalent
to M-theory on an AdS soliton with $p=5$  in a direct product with a four-sphere.
The AdS soliton geometry is
\beqa
ds^2 &=& \frac{r^2}{L^2}\left(f(r)d\tau^2 + dz^2 +
\eta_{\mu\nu}dx^\mu dx^\nu\right)+
\frac{L^2}{r^2}f^{-1}(r)dr^2   
\nonumber
\\
&& {\rm with}\ f(r) =1-\frac{R^{6}}{r^{6}}
\label{eq:elevend}
\eeqa
where $z$ and $\tau$ are the coordinates parameterizing the first and second
circles above, respectively. As described above,
dimensionally reducing on $z$ yields Type IIa string theory 
on the throat geometry of a(n appropriate) D4-brane.
Now with respect to the $x^\mu$-directions, there are two spin-zero modes
in the metric fluctuations. The first would be the polarization identified
as the scalar mode given in eq.~\reef{guess} with $p=5$. The second would be a
mode identified as a transverse traceless mode in section 3.1 with 
a polarization tensor of the form
\beq
\veps_{zz}=-4\quad{\rm and}\quad\veps_{\mu\nu}=\eta_{\mu\nu}+{k_\mu k_\nu\over M^2}\ .
\labell{skald}
\eeq
As both of these modes would contain $h_{zz}$ fluctuations, they would both
seem to mix with the ten-dimensional string theory dilaton \cite{berg}.
We are not sure how these two modes are distinguished, but
presumably one should decouple while the other is dual to spin-zero
glueballs associated with the operator Tr$F^2$ in QCD$_4$.
In any event, at least with in the WKB approximation, the spectra
of both modes are degenerate, and so our calculations suggest a degeneracy
of the spin-two and spin-zero glueballs at least in the large $N$ limit
of QCD$_4$.

\vspace{1cm}
{\bf Acknowledgements}

This research was supported by NSERC of Canada and Fonds
FCAR du Qu\'ebec. 
We would like to acknowledge useful conversations with Gary Horowitz,
Guy Moore and Soo-Jong Rey. RCM would also like to thank the Institute for Theoretical
Physics at UCSB for its hospitality in the final stages of this work.
While at the ITP, RCM was supported by NSF Grant PHY94-07194.

We have been informed that S. Lee, J. Park, S. Moon
 and S.-J. Rey \cite{prep} have also considered
the graviton spectra for general D$p$-brane backgrounds, and had observed
the degeneracy of the spectra for the dilaton and transverse traceless
gravitons on the AdS$_5$ soliton. 

\appendix
\section{Appendix}

In this appendix, we provide some of the details for the analysis of the 
metric fluctuations with nontrivial $\tau$-dependence. In this case the ansatz
made in section 3 is extended to $h_{ab}=H_{ab}(r)e^{i(k\cdot x+q\tau)}$
where $q$ is the momentum 
in the compact direction given by  $q=2\pi \tn/\beta$ for any integer $\tn$ 
and the period $\beta$ is given in eq.~\reef{period}.  
For the purposes of gauge fixing, it is natural to include $q$ as the last
component of a $(p+1)$-dimensional momentum vector, $k^a=(k^\mu,q)$.
For the sake of simplicity,
we will choose to work in the $p$-dimensional ``rest frame'' of these excitations, \ie
with an appropriate Lorentz boost, we set $k^a=(M,0,\ldots,0,q)$.
Now the natural transversality condition to impose on the metric perturbations
becomes
\beq
H_{ab}k^{b}=0 \Rightarrow H_{at}=-\frac{q}{M}H_{a\tau}\ .
\labell{qtrans}
\eeq
We begin with extension of the simple transverse traceless polarizations. As in
eq.~\reef{ttpol}, we set
\beq
H_{ab}=\varepsilon_{ab}\,{r^2\over L^2}H(r)
\labell{ttpola}
\eeq
where $\veps_{\tau a}  = \veps_{r a}  =  \veps_{t a} = 0$.
Substituting the extended ansatz into eq.~\reef{eq:lingrav}, one finds that the only 
effect of adding the compact momenta is to shift the mass term in
eq.~\reef{ttode} by $M^2\rightarrow M^2-q^2/f(r)$ where $f(r)$ is the function
appearing in the metric
in eq.~\reef{eq:metric}. After performing the same wavefunction redefinition and
change of variables as described in section 3.1, the equation of motion
becomes Schr\"odinger-like \reef{ttschro} with an effective potential
\beqa
V(y) & =& \frac{1}{4} +\frac{e^{2y}\left(
p(p+2)\left(1+e^y\right)^{2(p+1)}
-2p(p+2)\left(1+e^y\right)^{p+1}-1\right)}{4\left(1+e^y\right)^2\left(
\left(1+e^y\right)^{p+1}-1\right)^2}
\nonumber
\\
&&\ -\frac{M^2L^4}{R^2}\frac{e^{2y}\left(1+e^y\right)^{p-3}
}{\left(1+e^y\right)^{p+1}-1}
+\frac{q^2L^4}{R^2}\frac{e^{2y}
\left(1+e^y\right)^{2p-2}}{\left((1+e^y)^{p+1}-1\right)^2}\ .
\labell{eq:ttpot2}
\eeqa
The latter has the following asymptotic behavior
\beqa
V(y \gg 0)&=&{(p+1)^2\over4}-\frac{p(p+2)}{2}e^{-y}
+\left({p(p+2)\over4}-{M^2L^4\over R^2}\right)e^{-2y}
\nonumber
\\
V(y\ll 0)&=&\frac{\tn^2}{4}+\left({p+2\over4}-\frac{M^2L^4}{(p+1)R^2}+
\frac{\tn^2}{4}\right)e^y .
\labell{asympot2}
\eeqa
Note that the turning point for $y\ll 0$ is now at a finite coordinate distance
due to the inclusion of a non-zero compact momenta. This has the anticipated effect
of shifting the mass spectrum upwards from that given in eq.~\reef{eq:ttspec}.
In any event, one can verify that for $M^2<0$ the potential is everywhere positive
and so the Schr\"odinger equation yields no normalizable zero-energy solutions.
Hence, since the spectrum has $M^2>0$ for all of these spin-two modes, they clearly do not 
present any difficulty for the stability of the AdS soliton.
Finally, we also note that the above results are again identical to that 
for the massless, minimally coupled scalar in the case that internal momentum is included.
The the spectra of these two set of modes remains identical when one allows for
nontrivial $\tau$-dependence.

Next we consider the vector modes with nontrivial $\tau$-dependence.
In this case, the ansatz for the polarization of eq.~\reef{eq:eppol} 
had to be modified as well. Using the transverse gauge condition \reef{qtrans}
in the rest frame, a consistent ansatz is given by
\beq
H_{\tau i}={r^2\over L^2}H(r), \quad H_{ti}=-\frac{q}{M}\frac{r^2}{L^2}H(r),
\quad H_{ri}=-\frac{q}{M}\frac{r^2}{L^2}a(r)
\labell{qpol}
\eeq
for some $i\in\lbrace1,\ldots,p-1\rbrace$, and all other components of
$H_{ab}$ are zero. Consistency of this ansatz in the equations of
motion~\reef{eq:lingrav} requires that
\beq
a(r)=\frac{R^{p+1}}{r^{p+1}\left(1+q^2/M^2\right)-R^{p+1}}
{1\over M}\frac{\partial H(r)}{\partial r}
\labell{ggg}
\eeq
Redefining the radial profile according to $H(r)=\alpha(r)\phi(r)$ with
\beq
\alpha=\sqrt{{r-R\over r^{p+2}}\,{r^{p+1}\left(1+q^2/M^2\right)-R^{p+1}
\over r^{p+1}-R^{p+1}}}
\labell{exalph}
\eeq
and then making the usual change of variables, $r=R(1+e^y)$,
leads to an effective potential of the form,
\beqa
V(y)& &= \frac{1}{4}+\frac{p(p+2)e^{2y}}{4\left(1+e^y\right)^2}-\frac{M^2L^4}
{R^2}\frac{e^{2y}\left(1+e^y\right)^{p-3}
\left(\left(1+e^y\right)^{p+1}\left(1+q^2/M^2\right)-1\right)}
{\left((1+e^y)^{p+1}-1\right)^2}
\nonumber
\\
&&-\frac{q^2}{M^2}\frac{p(p+2)\left(\left(1+e^y\right)^{p+1}(2+q^2/M^2)
-2(3+q^2/M^2)\right)\left(1+e^y\right)^{3p+3}e^{2y}}{4\left(1+e^y\right)^2
\left(\left(1+e^y\right)^{p+1}-1\right)^2\left(\left(1+
e^y\right)^{p+1}\left(1+q^2/M^2\right)-1\right)^2}
\nonumber
\\
&&+\frac{q^2}{M^2}\frac{\left(1+e^y\right)^{2p+2}
(4+q^2/M^2)+4(1+e^y)^{p+1}e^{2y}}
{4\left(1+e^y\right)^2\left(\left(1+e^y\right)^{p+1}-1\right)^2\left(\left(1+
e^y\right)^{p+1}\left(1+q^2/M^2\right)-1\right)^2}
\labell{qpot}
\eeqa
Note that this is nearly identical to eq.~\reef{eq:eepot} except for the
last term which is proportional to $q^2$.
This effective potential has the following asymptotic behavior,
\beqa
V(y \gg 0)&=&{(p+1)^2\over4}-\frac{p(p+2)}{2}e^{-y}
+\left({p(p+2)\over4}-{M^2L^4\over R^2}(1+q^2/M^2)\right)e^{-2y}
\nonumber 
\\
V(y\ll 0)&=& \frac{\tn^2}{4}+\left(\frac{(p+2)\left(q^2-M^2\right)-pM^2}
{4q^2}-\frac{M^2L^4}{R^2}\frac{p+1-(p-2)q^2/M^2}{(p+1)^2}\right)e^y
\labell{exqpot}
\eeqa
Note that setting $q=0$ in the expression for $V(y\ll 0)$ is not consistent. 
This simply reflects the fact that the two limits $y\rightarrow -\infty$ and 
$q\rightarrow 0$ do not commute. In any event,
comparing these limiting values to those of 
eq.~\reef{exasymp} one sees in a $\frac{1}{M}$ expansion that a non-vanishing 
compact momenta will lead to a larger mass gap than in the $q=0$ case.
Alternatively, one can show that the potential is positive everywhere for
$M^2<0$, and hence there are no tachyonic excitations. 
That is the AdS soliton is perturbatively stable against these vector modes
with nontrivial $\tau$-dependence.

Finally, we consider the scalar modes for which the ansatz for the polarization must
be extended beyond that in eq.~\reef{guess}. Again,
the transversality constraint \reef{qtrans} and consistency
of the equations of motion guide us in constructing the following ansatz
\beqa
H_{\tau\tau}(r)&=&-{r^2\over L^2}f(r)H(r)
\nonumber \\
H_{ij}(r)&=&\frac{1}{p-1}{r^2\over L^2}H(r)\,\delta_{ij} \;\;\;i,j=1,...,p-1 
\nonumber \\
H_{tt}(r)&=&\frac{q^2}{M^2}H_{\tau\tau}(r)
\nonumber \\
H_{rr}(r)&=&\frac{L^2}{r^2}f^{-1}(r)c(r)
\nonumber\\
H_{rt}(r)&=&-iM\,d(r)
\nonumber \\
H_{\tau t}(r)&=&-\frac{q}{M}H_{\tau\tau}(r)
\nonumber \\
H_{\tau r}(r)&=&iq\,e(r)
\nonumber \\
\labell{qguess}
\eeqa
where $f(r)$ is the function appearing in the metric \reef{eq:metric}, while
$c(r),d(r)$ and $e(r)$ are functions fixed by consistency of the equations
of motion \reef{eq:lingrav}. When $q=0$, this ansatz corresponds to the
polarization in eq.~\reef{guess} with $b(r)=0$.
Substituting the ansatz into the linearized Einstein equations \reef{eq:lingrav},
one can eliminate
the functions $c(r),d(r),e(r)$ in terms of $H(r)$ and its derivatives
to obtain a single second order equation for $H(r)$. Next one redefines the radial
profile by $H(r)=\eta(r)\psi(r)$ with
\beq
\eta(r)=\frac{\left(2p\left(1-q^2/M^2\right)r^{p+1}-(p-1)R^{p+1}\right)
\sqrt{\frac{r-R}{r(r^{p+1}-
R^{p+1})}}}
{\left(2r^{p+1}\left(p-(p-1)q^2/M^2\right)\left(1-q^2/M^2\right)-(p-1)R^{p+1}\left(
1-4q^2/M^2+q^4/M^4\right)\right)}
\labell{qdiagdef}
\eeq
%nn signs on q^2/M^2 terms may need fixing???
%r I'm looking into that.
%NNN signs here needed to get fixed right?
and changes to the $y$ coordinate as before.  In the end, $\psi(r)$ satisfies
the Schr\"odinger equation with the effective potential
\beqa
V(y)& =& \frac{1}{4}+ \frac{e^{2y}\left(
p(p+2)\left(1+e^y\right)^{2(p+1)}
-2p(p+2)\left(1+e^y\right)^{p+1}-1\right)}{4\left(1+e^y\right)^2\left(
\left(1+e^y\right)^{p+1}-1\right)^2}
\nonumber \\
&&-\frac{L^4M^2}{R^2}\frac{\left(\left(1+e^y\right)^{p+1}
(1+q^2/M^2)-
1\right)\left(1+e^y\right)^{p-3}e^{2y}}{\left((1+e^y)^{p+1}-1\right)^2}
\nonumber \\
&&-{2(p-1)(p+1)^3\left(1+e^y\right)^{p-1}e^{2y}\over
\left(\left(1+e^y\right)^{p+1}-1\right)
\left(2p\left(1+e^y\right)^{p+1}(1+q^2/M^2)-p+1\right)^2}
\nonumber
\\
&&-\frac{4q^2}{M^2}\frac{e^{2y}\left(p\left(2p-1\right)\left(1
+e^y\right)^{2p+2}
+p\left(p-1\right)\left(p+1\right)^2\left(1+e^y\right)^{p+1}\right)}{\left(1+
e^y\right)^2\left(\left(1+e^y\right)^{p+1}-1\right)\left(2p\left(1+e^y\right)
^{p+1}(1+q^2/M^2)-p+1\right)^2}
\labell{diagqpot}
\eeqa
One can easily see that setting $q=0$ in this expression reproduces the result
in eq.~\reef{eq:diagpot}.
This potential behaves in the asymptotic regions as
\beqa
V(y \gg 0)&=&{(p+1)^2\over4}-\frac{p(p+2)}{2}e^{-y}
+\left({3p(p+2)\over4}-{M^2L^4\over R^2}(1+q^2/M^2)\right)e^{-2y}
\nonumber 
\\
V(y\ll 0)&=&\left(\frac{2p(p+2)q^2+(7p-10)(p+1)M^2}{2pq^2-(p+1)M^2}+
\frac{(p-2)\tn^2}{4}-\frac{M^2L^4}{(p+1)R^2}\right)e^y
\labell{diaglims}
\eeqa
Note that in this case, the two limits $y\rightarrow -\infty$ and 
$q\rightarrow 0$ do commute for the present calculations. 
Once again it can be seen that to leading order in $\frac{1}{M}$ the potential
near the interior turning point is increased by the inclusion of a 
non-zero value of the compact momentum however the turning point remains at
$y=-\infty$. Further one can show again that there are no tachyonic excitations
since the potential is everywhere positive for $M^2<0$.
Hence as expected, these scalar modes do not yield any instabilities for the AdS soliton
when one allows for a nontrivial $\tau$-dependence.

\end{document}